\documentclass[a4paper,11pt]{article}
\usepackage{graphicx,amssymb,amsmath,bm,latexsym,color,epsf,cite,comment}
\makeatletter
\@addtoreset{equation}{section}

\makeatother
\newcommand{\bea}{\begin{eqnarray}}
\newcommand{\eea}{\end{eqnarray}}
\newcommand{\vs}[1]{\vspace{#1 mm}}
\newcommand{\hs}[1]{\hspace{#1 mm}}
\renewcommand{\a}{\alpha}
\renewcommand{\b}{\beta}

\renewcommand{\d}{\delta}
\newcommand{\e}{\epsilon}

\newcommand{\s}{\sigma}

\newcommand{\G}{\Gamma}
\newcommand{\vp}{\varphi}
\newcommand{\la}{\lambda}
\newcommand{\pa}{\partial}
\newcommand{\nn}{\nonumber\\}

\newcommand{\lan}{\langle}
\newcommand{\ran}{\rangle}

\newcommand{\tg}{\tilde g}
\newcommand{\hg}{{\hat g}}

\newcommand{\Tr}{{\rm Tr}}

\begin{document}

\begin{flushright}
NITEP 266
\end{flushright}
\begin{center}
{\large\bf Noncritical Conformal Gravity and Four-Dimensional Liouville Theory}
\vs{10}

{\large
Hikaru Kawai$^{a,b,c,}$\footnote{e-mail address: hikarukawai@phys.ntu.edu.tw}
and
Nobuyoshi Ohta$^{c,d,}$\footnote{e-mail address: ohtan.gm@gmail.com}
} \\
\vs{5}

$^a${\em Department of Physics and Center for Theoretical Physics, National Taiwan University, Taipei 106, Taiwan}
\vs{1}

$^b${\em Physics Division, National Center for Theoretical Sciences, Taipei 106, Taiwan}
\vs{2}

$^c${\em Nambu Yoichiro Institute of Theoretical and Experimental Physics (NITEP), \\
Osaka Metropolitan University, Osaka 558-8585, Japan}
\vs{1}

$^d${\em Institute of Fundamental Physics and Quantum Technology,
Department of Physics, School of Physical Science and Technology,
Ningbo University, Ningbo, Zhejiang 315211, China}

\vs{10}
%%%%%%%%%%%%%%%%%%%%%%%%%%%%%%%%
{\bf Abstract}
\end{center}

We study the quantum aspects of the conformal gravity in four dimensions, specifically addressing a known discrepancy
in beta functions between general quadratic curvature theories and conformal gravity, which corresponds to two scalar
degrees of freedom. We demonstrate that this mismatch is resolved by carefully introducing gauge-fixing and
ghost terms via the BRST symmetry, which effectively adds the two scalar modes. Drawing lessons from two-dimensional
quantum gravity and Liouville theory, we proceed to integrate the four-dimensional trace anomaly to
derive a consistent Liouville action, which is given by a free-field action for the conformal mode
with a consistent conformal anomaly.
Finally we give the condition that the BRST transformation is anomaly free.

\setcounter{footnote}{0}

%%%%%%%%%%%%%%%%%%%%%%%%%%%%
\section{Introduction}
%%%%%%%%%%%%%%%%%%%%%%%%%%%%

In this paper we study the quantization of conformal gravity in four dimensions.
The conformal sector is described by the conformal factor $e^{\phi(x)}$ in spacetime metric of the form
\bea
g_{\mu\nu}(x)=e^{\phi(x)} \hg_{\mu\nu}(x),
\label{conffactor}
\eea
where $\hg_{\mu\nu}$ is a family of reference metric representing the degrees of freedom apart from
the conformal mode.
Before going into this subject, let us first briefly summarize a puzzle in the beta functions for quadratic
curvature theory, which helps us to better understand the theory.

It has been known for some time that there is a subtlety in the beta functions of the general quadratic curvature
theory
\bea
S=\int d^4 x \sqrt{g}\, \left( \frac{1}{2\la} C_{\mu\nu\a\b}^2
+\frac{1}{\xi} R^2 - \frac{1}{\rho} R_{\rm GB}^2\right),
\label{generalaction}
\eea
and the conformal gravity
\bea
S=\int d^4 x \sqrt{g}\, \left( \frac{1}{2\la} C_{\mu\nu\a\b}^2 - \frac{1}{\rho} R_{\rm GB}^2\right),
\label{confaction}
\eea
where $\la,\xi$ and $\rho$ are dimensionless couplings, and $C_{\mu\nu\a\b}$ is the Weyl tensor which is
defined such that its trace is zero. Its square is given as
\bea
C_{\mu\nu\la\s}^2 = R_{\mu\nu\la\s}^2 - 2 R_{\mu\nu}^2+\frac{1}{3} R^2,
\label{confg}
\eea
and $R_{\rm GB}^2$ is the Gauss-Bonnet (GB) term
\bea
R_{\rm GB}^2 = R_{\mu\nu\a\b}^2 - 4 R_{\mu\nu}^2 +R^2,
\label{GBt}
\eea
Tbe beta functions in this system may be calculated by the functional renormalization group equation
\bea
k \frac{d}{dk}\G_k=\frac12 \Tr \left(\G_k^{(2)}+R_k\right)^{-1} k\frac{d}{dk} R_k,
\label{flow}
\eea
where $\G_k$ is the effective average action, $k$ is the cutoff, $R_k$ is a cutoff function suppressing
the contribution of the modes below the momentum scale $k$ to $\G_k$, and $\G_k^{(2)}$ is the second variation
of the effective average action.
The right-hand side of \eqref{flow} gives the anomaly coeffectients $a,b$, and $c$:
\bea
\int d^4 x \sqrt{g} \frac{1}{(4\pi)^2} \left( c\, C_{\mu\nu\la\s}^2 + b R^2 - a\, R_{\rm GB}^2  \right).
\label{abccoeff}
\eea
in terms of which the beta functions are given as
\bea
\beta_\la &=& - \frac{2c}{(4\pi)^2} \la^2, \nn
\beta_\xi &=& - \frac{b}{(4\pi)^2} \xi^2, \nn
\beta_\rho &=& - \frac{a}{(4\pi)^2} \rho^2.
\eea

The beta functions for the general quadratic curvature theory have been calculated in \cite{FT,AB,BS2,OP},
while those for the conformal gravity have been calculated in~\cite{FT,BS1,OP2}.
Naively it is expected that the beta functions for the conformal gravity are reproduced if we take
the limit $\xi\to\infty$~\cite{SS,KO}. Surprizingly enough, it is found that they do not agree.
Rather there are discrepancy corresponding to precisely two scalar degrees of freedom.
This may arise because the conformal gravity, to which the general quadratic theory appears to reduce
in the $\xi\to\infty$ limit,  has additional gauge symmetry under the conformal transformation;
in this limit, the conformal symmetry is recovered, so without fixing it, the theories
may not coincide. The problem is recently understood by making partial gauge fixing in the conformal
gravity~\cite{MPSVZ}. We then discuss the resolution, and find the formulation can be used to derive
the consistent quantum conformal gravity with four-dimensional Liouville theory.

This paper is organized as follows.
In Sect.~\ref{conf}, based on the BRS transformation for the conformal invariance, we fix the degrees
of freedom of the conformal modes and evaluate the beta functions.  Indeed, we can confirm that
the discrepancy mentioned above can be explained as the contribution of the ghost and anti-ghost fields
for this gauge fixing.
Section \ref{der} reviews the derivation of the Liouville action in two dimensions and the conditions
for the correct quantization of the conformal modes.
Using this as a reference, Sect.~\ref{4D} derives the Liouville action for the four-dimensional case and
provides consistent quantization conditions for the conformal modes. In the two-dimensional case,
the quantization condition was that the total central charge involving the Liouville field vanishes.
Similarly, for the four-dimensional case, we can confirm that this condition is replaced by the requirement
that the total beta function involving the Liouville field and the ghosts of the conformal transformation vanishes.
We show that this uniquely leads to a free field action for the conformal mode.
We also contrast this approach with the widely-used Riegert proposal~\cite{R} for the four-dimensional
Liouville action, providing a more robust, consistent formulation derived through the quantization procedure.
We summarize our results in Sect.~\ref{summary}.
Some necessary formulae and discussions are relegated to the appendices. In Appendix~\ref{conft}, we display
the transformation property of the curvaturs under the conformal transformation.
In Appendix~\ref{confs}, we discuss the transformation properties of scalar field under the conformal transformation.
Finally in Appendix~\ref{consistency}, we summarize the consistency condition for the trace anomaly.

\section{Conformal anomaly from the conformal gravity}
\label{conf}

Let us consider the action for the conformal gravity~\eqref{confaction}.
Since this has the conformal symmetry under the Weyl transformation $g_{\mu\nu}\to e^\s g_{\mu\nu}$,
to quantize the system, we have to gauge fix it.
To identify the gauge fixing and the corresponding Faddeev-Popov ghost terms, we use the BRST symmetry.
The infinitesimal conformal or more precisely Weyl transformation is given as
\bea
\d g_{\mu\nu} = \e g_{\mu\nu},
\eea
where $\e$ is an infinitesimal transformation parameter. This is promoted to the BRST transformation
\bea
\d_B g_{\mu\nu}= \d\la\, c g_{\mu\nu},
\label{brst1}
\eea
where $\d\la$ is an anticommuting parameter and $c$ is a ghost field for the conformal
transformation.\footnote{
This ghost $c$ and the following antighost $b$ should not be confused with the $a,b,c$ coefficients
in Eq.~\eqref{abccoeff}.}
The BRST transformation of the ghost is determined by the requirement that the BRST transformation~\eqref{brst1}
be nilpotent. This gives
\bea
\d_B c=0.
\label{brstc}
\eea
Finally the BRST transformations of the antighost $b$ and the associated auxiliary field $B$ would be
naively defined as
\bea
\d_B b \stackrel{?}{=} \d\la B, \qquad
\d_B B \stackrel{?}{=} 0.
\label{gtrans}
\eea
However we expect that the ghosts $(b,c)$ have the conformal weight $(-1,0)$.
The transformation~\eqref{gtrans} does not take into account the conformal weight of the antighost $b$.
It turns out~\cite{MPSVZ} that we can incorporate the conformal weight by the following modifications
\bea
\d_B b = \d\la (-cb+B),\qquad
\d_B B = -\d\la cB,
\label{brstb}
\eea
which are still nilpotent.
Under this transformation, we find from the result in Appendix~\ref{conft} that
the determinant of the metric and the scalar curvature transform as
\bea
\d_B \sqrt{g} = 2 \d\la c \sqrt{g}, \qquad
\d_B R = -\d\la ( c R +3 \square c).
\eea
We introduce $R$ as a partial gauge fixing function of the conformal symmetry:
\bea
f=R.
\label{partialg}
\eea
Conformal invariant quantities can be computed integrating over an arbitrary smearing of the gauge condition
around $f=0$. The gauge fixing term and the Faddeev-Popov ghost term may then be written as~\cite{KU,ON}
\bea
{\cal L}_{\rm GF+FP}\hs{-2}&=&\hs{-2} \d_B \left[ \sqrt{g} b \left(R-\frac{\a}{2} B\right) \right]/\d\la \nn
\hs{-2}&=&\hs{-2} \sqrt{g} \left[ 2c b\left(R-\frac{\a}{2} B \right) +(-cb+B) \left(R-\frac{\a}{2}B \right)
+ b(cR + 3 \square c) -\frac{\a}{2} bc B \right],\qquad
\eea
where $\a$ is a gauge parameter.
This gives the action
\bea
S_{\rm GF+FP} = \int d^4 x \sqrt{g} \left[ -\frac{\a}{2}\left( B -\frac{1}{\a} R \right)^2
+\frac{1}{2\a} R^2 + 3 b \square c \right],
\label{actiongg}
\eea
where anticommuting property of the ghosts is used so that $bc=-cb$.\footnote{
If we use the transformation~\eqref{gtrans}, the same procedure gives
$$
S_{\rm GF+FP} = \int d^4 x \sqrt{g} \left[ -\frac{\a}{2}\left( B +cb -\frac{1}{\a} R \right)^2
+\frac{1}{2\a} R^2 + 3 b \square c \right].
$$
So practically both transformations give the same result. The original conformal transformation is broken
by the gauge fixing term, leaving only the global BRST invariance. Thus we may adopt transformations
for the antighost and auxiliary fields different from those suggested by the conformal transformation
as long as we consider the proper transformation for the other physical fields.}
It is interesting that there is no $Rbc$ term in the action~\eqref{actiongg} even though
$b$ and $c$ are (fermionic) scalar fields, in contrast to usual bosonic scalar fields (see Appendix~\ref{confs}).

Since the gauge parameter $\a$ is arbitrary, we can choose it as we wish.
We then find that the total gauge fixed action~$\eqref{confaction}+\eqref{actiongg}$ is,
apart from the gauge fixing and Faddeev-Popv terms, coincide the general quadratic gravity action~\eqref{generalaction}
if we set $2\a=\xi$.

After the gauge fixing of the reparametrization, by use of the heat kernel expansion, the rhs of this equation
was calculated as~\cite{FT,AB,BS2,OP}
\bea
\int d^4 x \sqrt{g} \frac{1}{(4\pi)^2} \left[ \frac{133}{20}C_{\mu\nu\la\s}^2
+ \frac{5(72 \la^2-36 \la\xi +\xi^2)}{36\, \xi^2} R^2
-  \frac{196}{45} R_{\rm GB}^2 \right],
\label{counter0}
\eea
where only the $b_4$ coefficients are kept.
The beta functions for the quadratic terms are then
\bea
\beta_\la^{\,\rm g} &=& -\frac{1}{(4\pi)^2} \frac{133}{10} \la^2, \nn
\label{gbeta}
\beta_\xi^{\,\rm g} &=& -\frac{1}{(4\pi)^2} \frac{5}{36}(72 \la^2-36 \la\xi +\xi^2), \\
\beta_\rho^{\,\rm g} &=& -\frac{1}{(4\pi)^2} \frac{196}{45} \rho^2. \nonumber
\eea
If we consider the limit of $\xi \to \infty$ in Eq.~\eqref{counter0}, we expect that this would give the result
for the conformal gravity without $R^2$ term, and the beta function for $\xi$ is absent.
However, the beta functions for the conformal gravity was calculated in several works~\cite{FT,BS1,OP2},
and the result is given as
\bea
\beta_\la^{\,\rm conf} &=& -\frac{1}{(4\pi)^2} \frac{199}{15} \la^2, \nn
\beta_\rho^{\rm conf} &=& -\frac{1}{(4\pi)^2} \frac{87}{20} \rho^2 ,
\label{cbeta}
\eea
which are different from the limit of $\xi\to\infty$ in \eqref{gbeta}.
This had been a mystery for some time.

The above result is then written in terms of $a,b,c$ coefficients in \eqref{abccoeff} as
\bea
c^{\,\rm g} =  \frac{133}{20},\quad
&& a^{\, \rm g} = \frac{196}{45}, \\
c^{\,\rm conf} =  \frac{199}{30},\quad
&& a^{\, \rm conf} =  \frac{87}{20},
\label{acconf}
\eea
Their difference is
\bea
c^{\,\rm g} -c^{\,\rm conf}=\frac{1}{60}, \nn
a^{\,\rm g} -a^{\,\rm conf}=\frac{1}{180}.
\label{mis}
\eea
On the other hand, it is known that the contribution of the matter is
\bea
c^{\rm m}=\frac{1}{120}(N_S + 6N_F +12 N_V),\nn
a^{\rm m}=\frac{1}{360}(N_S + 11N_F +62 N_V),
\label{acm}
\eea
where $N_S, N_F$ and $N_V$ are the number of scalar, fermion and vector fields, respectively.
It is then clear that there is mismatch in the coefficients~\eqref{mis} precisely by the contribution
of two scalar degrees of freedom~\cite{FT,SS,MPSVZ,KO}.
Interesting enough, the above formulation with gauge fixing function~\eqref{partialg} beautifully explains
the mismatch. We have seen that the gauge fixing procedure introduces precisely two scalar (fermionic) modes
$b$ and $c$.
This is the reason why there is an apparent mismatch of two scalar degrees of freedom.
The gauge fixing and ghost terms~\eqref{actiongg} actually decouple from the physical space.
This is consistent with the fact that the beta function for $\la$ is independent of $\xi$.

However, the gauge-fixed theory constructed as above cannot necessarily be regarded as a conformally
invariant gravitational theory. This is because the conformal invariance is broken unless the total beta
functions are zero (see~\eqref{cbeta}):
\bea
&& \frac{199}{30}+c^{\rm m}=0, \nn
&& \frac{87}{20}+a^{\rm m}=0.
\label{4Dcritical}
\eea

Generically, except in special cases satisfying the above conditions, even if a system including gravity
is classically conformally invariant, quantization breaks the symmetry due to the anomaly.
In this case, the conformal modes do not appear in the classical action but do appear in the effective action.
Therefore, the path integral for the conformal modes requires careful consideration.
This is a four-dimensional analogue of noncritical strings in two dimensions and may be called noncritical
conformal gravity.

In two dimensions, this problem has been resolved by Liouville theory.
To study this in four dimensions, let us start by reviewing the two-dimensional situation.

\section{Lessons from two-dimensional quantum gravity for conformal gravity -- Integrating the trace anomaly}
\label{der}

The action of two-dimensional gravity is
\bea
S_2 = \int d^2 \xi \sqrt{g} \left( R + \mu \right),
\eea
where we have set the coefficient of the Einstein term to be 1 and $\mu$ is the cosmological constant.
We use the Eucleadean metric.
The Einstein term in two dimensions gives just topological number
\bea
\int d^2\xi \sqrt{g} R = 4\pi \chi =\mbox{const.}
\eea

The energy-momentum tensor is given by the partition function $Z[g]$ as
\bea
\langle T_{\mu\nu} \rangle &=& -\frac{2}{\sqrt{g}} \frac{\d \log Z[g]}{\d g^{\mu\nu}}.
\eea
This implies that
\bea
\d \log Z[g] =-\int d^2 \xi \frac12 \sqrt{g}\, \d g^{\mu\nu} T_{\mu\nu}.
\eea
When the conformal matter is coupled, there appears trace anomaly.
In such a case, under the scale transformation $\d g^{\mu\nu}=-g^{\mu\nu} \d\phi$, we have the relation
\bea
\d(\log Z[g]) :=\log(Z[g e^{\d\phi}])-\log(Z[g]) = \int d^2 \xi\, \frac12\, \sqrt{g}\,\d\phi\, T^\mu_\mu
\eea
The trace anomaly is given by
\bea
\lan T^\mu_\mu \ran =\frac{c}{24\pi} R,
\eea
where $c$ is called the central charge.

Write the partition function as
\bea
Z=e^{-F},
\eea
and we have
\bea
\d F =F[g e^{\d\phi}]-F[g] =-\frac{c}{48\pi} \int d^2 \xi \sqrt{g} R \d\phi.
\label{difftrace}
\eea
Now we write $\d\phi=\phi dt$, and Eq.~\eqref{difftrace} is
\bea
dF=-\frac{c}{48\pi}\int d^2\xi \sqrt{g(t)} R(t) \phi dt,
\label{intgrat}
\eea
where $R(t)$ is the curvature for $g(t)=g e^{t \phi}$. Using the formula~\eqref{scalart}
and integrating \eqref{intgrat} from $t=0$ to $t=1$, we obtain
\begin{align}
F[g e^{\phi}]-F[g] &=-\int_0^1 \frac{c}{48\pi} \int d^2\xi \sqrt{g} (R-t \square \phi)\phi dt \nn
&=-\frac{c}{48\pi}\int d^2 \xi \sqrt{g}\left(\frac12 g^{\mu\nu} \pa_\mu\phi \pa_\nu \phi+R\phi \right),
% =: S_L[g,\phi],
\label{liouville}
\end{align}
which is proportional to the Liouville action.

The same result can be obtained as follows. The renormalized partition function is
written by the bare partition function with counterterm
\bea
Z[g] = \lim_{\e\to 0} Z[g_0] e^{\frac{c}{48\pi \e_2}\int d^2\xi\sqrt{g}\, R},
\label{counter1}
\eea
where $\e_2 =(2-D)/2$ and $c$ is a constant.
The Liouville action arises from the counterterm as the finite part of the counterterm.
Consider the Weyl transformation
\bea
g_{\mu\nu} = e^\phi \hg_{\mu\nu}.
\eea
From the formula~\eqref{scalart} given in Appendix~\ref{conft}, the counterterm in~\eqref{counter1} transforms into
\bea
-\frac{c}{48\pi \e_2} \int d ^D x \sqrt{\hg}\, e^{(D-2)\phi/2}
\left( \hat R -(D-1)\hat\square \phi -\frac{(D-1)(D-2)}{4} (\pa_\mu \phi)^2 \right),
\label{counter2}
\eea
where the D'Lambertian and covariant derivatives on the rhs are made of the hatted metric $\hg_{\mu\nu}$.
Similar notation should be understood in what follows.
The simple pole in the limit of $D\to 2 \ (\e_2 \to 0)$ is the genuine counterterm.
There are terms remaining finite in this limit.
We can find these terms by expanding the integrand in \eqref{counter2} in $(D-2)$ and keep the linear term.
This gives
\bea
&& \frac{c}{48\pi} \int d^2 \xi \sqrt{\hg}\left[ \phi \big( \hat R -\hat\square \phi \big)
-2 \hat\square\phi -\frac12 (\pa_\mu\phi)^2 \right] \nn
&&=\frac{c}{48\pi} \int d^2 \xi \sqrt{\hg} \left[ \frac{1}{2} (\pa_\mu \phi)^2 + \hat R \phi \right],
\eea
after partial integration, in agreement with Eq.~\eqref{liouville}.

The contribution of this Liouville action to the anomaly is given by
\bea
c_L=1-c.
\label{2DcL}
\eea
Here
1 is the quantum contribution whereas $-c$ is the classical contribution coming from the shift of $\phi$ under
the conformal transformation.
The absence of the total conformal anomaly determines $c$~\cite{D,DK}.
To show it more explicitly, let us write the contribution to the central charge from the matter as $c^{\rm m}$.
The contribution from the FP determinant or ghosts is $-26$, and that from the above Liouville action is $1-c$

Therefore, to have a consistent theory, these must add up to zero:
\bea
1-c -26 +c^{\rm m}=0.
\label{ctotal}
\eea
This assures the consistency of the theory. That is, the theory remains unchanged even if $\hat{g}$ is replaced
with $\hat{g}e^{\sigma}$.

Let us confirm the difference between critical and noncritical strings here. In critical strings, the total
central charge of the matter field and gravity is zero, and the conformal modes do not appear in either
the classical action or the quantum effective action. Therefore, we can simply forget about the conformal modes
by regarding them as gauge degrees of freedom. On the other hand, in noncritical strings, the conformal modes
do not appear in the classical action but do appear in the quantum effective action. The condition for
consistently quantizing this effective action is \eqref{ctotal}.

We will apply this method to the four-dimensional conformal gravity to identify the four-dimensional Liouville theory.

\section{Four-dimensional conformal anomaly and noncritical conformal gravity}
\label{4D}

\subsection{Integrating trace anomaly}
\label{integrate}

Using the method described in Sect.~\ref{der}, here we derive the four-dimensional Liouvill action.
Let us write
\bea
\langle T^\mu{}_\mu \rangle = \frac{1}{(4\pi)^2} \left[ c\, C_{\mu\nu\la\phi}^2
-a R_{\rm GB}^2 +b \square R \right].
\label{trace_anomaly}
\eea
We consider the dimensional regularization and try to calculate the anomaly terms
under the conformal transformation
\bea
g_{\mu\nu} = e^{\phi} \hg_{\mu\nu}.
\eea
The Weyl tensor squared in $D$ dimensions is
\bea
C_{\mu\nu\la\s}^2 = R_{\mu\nu\la\s}^2 - \frac{4}{D-2}R_{\mu\nu}^2+\frac{2}{(D-1)(D-2)} R^2,
\label{confgd}
\eea
and, according to the formulae in Appendix~\ref{conft}, it transforms as
\bea
\sqrt{g}\, C_{\mu\nu\la\s}^2 \to \sqrt{\hg}\, e^{((D-4)\phi/2} \hat C_{\mu\nu\la\rho}^2.
\eea
Hence the counterterm
\bea
\frac{c}{(4\pi)^2 \e}\int d^D x \sqrt{g}\, C_{\mu\nu\la\s}^2,
\eea
where $\e=(4-D)/2$, produces finite terms, in the limit of $D\to 4$, as
\bea
-\frac{c}{(4\pi)^2} \int d^D x \sqrt{\hg}\, \phi\, \hat C_{\mu\nu\la\s}^2.
\eea
%{\red Here $c$ is the matter and gravity contribution to the conformal anomaly.}

Similarly the GB term transforms as
\bea
\sqrt{g} R_{\rm GB}^2 &=& \sqrt{\hg}\, e^{(D-4)\phi/2} \Big[\hat R_{\rm GB}^2
+ 2(D-3) \hat R^{\mu\nu}(2 \hat\nabla_\mu \hat\nabla_\nu \phi-\pa_\mu\phi \pa_\nu\phi) -2 (D-3)\hat R\,
 \hat\square\phi \nn
&& -\frac{(D-3)(D-4)}{2}\hat R\, (\pa_\mu\phi)^2 -(D-2)(D-3) (\hat\nabla_\mu\hat\nabla_\nu\phi)^2 \nn
&& +(D-2)(D-3)(\hat\square\phi)^2 +(D-2)(D-3) (\hat\nabla_\mu\hat\nabla_\nu\s)\, \pa^\mu\phi\, \pa^\nu\phi \nn
&& +\frac{(D-2)(D-3)^2}{2}\hat\square\phi(\pa_\mu\phi)^2
+\frac{(D-1)(D-2)(D-3)(D-4)}{16}(\pa_\mu\phi)^2 (\pa_\nu\phi)^2\Big]. \nn
\label{gbconforig}
\eea
It is easy to make partial integration to get
\bea
\int d^D x \sqrt{g} R_{\rm GB}^2 \hs{-2}&=&\hs{-2} \int d^D x \sqrt{\hg}\, e^{(D-4)\phi/2}
\Big[\hat R_{\rm GB}^2 -(D-3)(D-4) \hat G^{\mu\nu} \pa_\mu\phi \pa_\nu\phi \nn
&& +\frac{(D-2)(D-3)(D-4)}{2}(\hat\nabla^\mu \hat\nabla^\nu \phi) \pa_\mu\phi \pa_\nu \phi \nn
&& +\frac{(D-2)(D-3)(D-4)(D-5)}{16} \{(\pa_\mu\phi)^2 \}^2\Big],
\label{gbconf}
\eea
where
\bea
\hat G^{\mu\nu} = \hat R^{\mu\nu} - \frac12 \hat R \hg^{\mu\nu},
\eea
is the Einstein tensor.
The terms depending on $\phi$ in Eq.~\eqref{gbconf} vanish for $D=4$, and this is consistent with the fact that
GB terms are total derivatives in $D=4$.
Hence the counterterm
\bea
- \frac{a}{(4\pi)^2 \e}\int d^D x \sqrt{g}\, R_{GB}^2,
\label{gbc}
\eea
gives
\bea
&& \frac{a}{(4\pi)^2 } \int d^4 x \Big[ \phi\, \hat R_{\rm GB}^2 -2 \hat G^{\mu\nu} \hat\nabla_\mu \phi
\hat\nabla_\nu \phi+ 2 (\hat\nabla^\mu \hat\nabla^\nu \phi) \pa_\mu\phi \pa_\nu\phi
- \frac14 \{(\pa_\mu\phi)^2 \}^2 \Big] \nn
&& = \frac{a}{(4\pi)^2}\int d^4 x \sqrt{g} \Big[ \phi\, \hat R_{\rm GB}^2 -2 \hat G^{\mu\nu} \hat\nabla_\mu \phi
 \hat\nabla_\nu \phi - (\pa_\mu\phi)^2 \hat\square \phi - \frac14 \{(\pa_\mu\phi)^2 \}^2 \Big].
\label{gbcr}
\eea
%{\red Here again $a$ is the matter and gravity contribution to the conformal anomaly.}

Collecting, our four-dimensional Liouville action would be
\bea
S_{\rm eff} \hs{-2}&=&\hs{-2} \int d^4 x \sqrt{\hg}\, \Big[ \frac{-c}{(4\pi)^2} \phi\, \hat C_{\mu\nu\la\s}^2
+ \frac{a}{(4\pi)^2} \Big( \phi\, \hat R_{\rm GB}^2
-2 \hat G^{\mu\nu} \pa_\mu \phi \pa_\nu \phi - (\hat\square\phi) (\pa_\mu \phi)^2
-\frac14 \{(\pa_\mu\phi)^2\}^2 \Big) \Big].\nn
\label{4dl}
\eea
Unfortunately this action contains nonlinear terms in the Liouville modes $\phi$ and is quite different
from the two-dimensional case.
Actually we should more carefully consider the quantization of the conformal gravity.
We will see that this leads to better action as discussed in the following.

\subsection{Four-dimensional noncritical conformal gravity}

We consider again the conformal gravity~\eqref{confaction} and write the partition function
\bea
Z=\int [{\cal D}g] Z_M[g] e^{-\int d^4 x \sqrt{g} (\frac{1}{2\la} C_{\mu\nu\la\rho}^2 -\frac{1}{\rho}R_{\rm GB}^2)}.
\eea
To quantize this theory, we should first gauge fix the conformal invariance.
For this purpose, we choose the conformal gauge fixing function~\eqref{partialg}.
As discussed in Sect.~\ref{conf}, this introduces the partial gauge fixing and ghost terms~\eqref{actiongg}.
The partition function takes the form
\bea
Z = \int [{\cal D}\!B {\cal D}\hg {\cal D}\vp{\cal D}b{\cal D}c] Z_M[\hg e^\vp] e^{-\int d^4 x \sqrt{g}
 (\frac{1}{2\la} C_{\mu\nu\la\rho}^2
-\frac{1}{\rho}R_{\rm GB}^2 +{\cal L}_{\rm GF+FP}) },
\eea
where
\bea
{\cal L}_{\rm GF+FP} = \frac{1}{2\a} R^2 + b \square c,
\eea
from Eq.~\eqref{actiongg}, where we have integrated out the auxiliary field $B$.
We then gauge fix the diffeomorphism and perform the path integral by using the background formalism:
\bea
Z= \int [{\cal D}\vp] Z_M[\hg] Z_{\rm G}[\hg] e^{- S_{\rm eff}[\hg,\vp]} ,
\label{GFFP}
\eea
where $Z_{\rm G}[\hg]$ is the gravity part for the background metric $\hg$.
In this formulation, any choice of $2\a=\xi$ is allowed since it is a gauge parameter.
We could also consider the contribution of the finite $R^2$ term to the Liouville action.
We will see that the consistency of the anomaly uniquely singles out its coefficient.
With this value, we find that the Liouville mode becomes free, while other choice gives nontrivial interacting theory.

Let us consider a combination of the finite term~\eqref{gbc} from the counterterm $R_{GB}^2$ and
the finite term $R^2$:
\bea
\left. -\frac{a}{(4\pi)^2 \e}\int d^D x \sqrt{g} R_{GB}^2 \right|_{\rm finite}
+ \frac{e}{(4\pi)^2} \int d^D x \sqrt{g} R^2 ,
\label{combi}
\eea
where we have chosen the coefficient of the first term such that it has the correct normalization to give
the corresponding term in \eqref{4dl}, and have written the coefficient of the second term as
$\frac{e}{(4\pi)^2}$ for convenience.
We use the method of Sect.~\ref{der} to extract the finite term as a Liouville action.
We would like to determine the coefficient $e$ by imposing the consistency discussed in Appendix~\ref{consistency}.
Under the conformal transformation $g_{\mu\nu} \to e^\phi \hg_{\mu\nu}$, we find from Eqs.~\eqref{combi}
and \eqref{q3}:
\bea
\int d^4 x \sqrt{\hg}\, \frac{1}{(4\pi)^2} \Big[ && \hs{-6}
a \Big\{ \phi\, \hat R_{\rm GB}^2 -2 \hat G^{\mu\nu} \hat\nabla_\mu \phi \hat\nabla_\nu \phi
- (\pa_\mu\phi)^2 \hat\square \phi - \frac14 \{(\pa_\mu\phi)^2 \}^2 \Big\} \nn
&& \hs{-6} + e\Big\{ \hat R^2 - 6 \hat R \hat\square\phi - 3 \hat R (\pa_\mu \phi)^2 + 9 (\hat\square\phi)^2
+ 9 \hat\square\phi (\pa_\mu\phi)^2 + \frac{9}{4} \{(\pa_\mu\phi)^2 \}^2 \Big\} \Big] \nn
= \int d^4 e \sqrt{\hg}\, \frac{1}{(4\pi)^2} \Big[ && \hs{-6}
\Big\{ a\, \phi\, \hat R_{\rm GB}^2 + e \hat R^2  - 6 e \phi \hat\square\hat R
- 2 a \hat G^{\mu\nu} \pa_\mu \phi \pa_\nu \phi  - 3e \hat R (\pa_\mu \phi)^2 + 9e (\hat\square\phi)^2 \nn
&& -(a-9e) (\pa_\mu\phi)^2 \hat\square \phi - \frac{(a-9e)}{4} \{(\pa_\mu\phi)^2 \}^2 \Big\}.
\label{intLiouville}
\eea
The conformal anomaly at the one-loop level gets contribution only from the quadratic terms in $\phi$.
From Eq.~\eqref{intLiouville}, they are given as
\bea
&& \int d^4 x \sqrt{\hg}\, \frac{9e}{(4\pi)^2} \Big[(\hat\square\phi)^2 - \frac{1}{3} \hat R (\pa_\mu \phi)^2
- \frac{2a}{9e} \hat G^{\mu\nu} \pa_\mu \phi \pa_\nu \phi \Big] \nn
&& = \int d^4 x \sqrt{\hg}\, \frac{9e}{(4\pi)^2} \phi \Big[\hat\square^2 + \Big\{ x \hat R^{\mu\nu}
+y \hat R \hg^{\mu\nu}\Big\} \hat\nabla_\mu \hat\nabla_\nu
+\frac13 (\pa_\mu \hat R) \pa^\mu \Big] \phi,
\label{squad}
\eea
Here in going from the first line to the next, we have made the partial integration, used the Bianchi identity
and set
\bea
x= \frac{2a}{9e}, \qquad
y= \frac13-\frac{x}{2}.
\eea
Using the formula from~\cite{Gu}, we can calculate their contribution to the anomaly:
\bea
&& \hs{-8} \frac{1}{(4\pi)^2}\frac{\G(2)}{2\G(3)} \left[ 2\Big(\frac{1}{90}\hat R_{\mu\nu\rho\phi}^2
- \frac{1}{90} \hat R_{\mu\nu}^2 +\frac{1}{36} \hat R^2 + \frac{1}{15} \hat\square \hat R \Big)
+\frac{2}{9}(x+4y)\hat\square \hat R -\frac{5}{18} (x+2y) \hat\square \hat R \right. \nn
&& \hs{-5}\left. + \frac{1}{24}(x+4y)^2 \hat R^2+\frac{1}{12}(x^2 \hat R_{\mu\nu}^2+(2xy+4 y^2) \hat R^2)
 +\frac{1}{6}(x+4y) \hat R^2-\frac13 (x \hat R_{\mu\nu}^2 +y \hat R^2)
+\frac{1}{3}\hat\square\hat R \right],\nn
\eea
which may be rewritten as
\bea
\frac{1}{1440(4\pi)^2} && \hs{-5} \left[ 3(5x^2-20x+4) \hat C_{\mu\nu\rho\phi}^2 -(15 x^2-60x+4)\hat R_{GB}^2
+25(x-2)^2 \hat R^2 \right. \nn
&& \left. +16(13-5x)\hat\square \hat R\right].
\label{anomalyf}
\eea
Requiring that there should be no $R^2$ term, as follows from the consistency condition discussed
in Appendix~\ref{consistency}, we find uniquely
\bea
x=2, \qquad
y=-\frac23,
\label{xy}
\eea
which in turn gives the condition
\bea
e=\frac{a}{9}.
\label{result1}
\eea
The action~\eqref{squad} reduces to
\bea
&& \int d^4 x \sqrt{\hat g}\, \frac{a}{(4\pi)^2} \phi \Big[\hat\square^2 + \Big\{ 2 \hat R^{\mu\nu}
-\frac23 \hat R \hat g^{\mu\nu}\Big\} \hat\nabla_\mu \hat\nabla_\nu
+\frac13 (\pa_\mu \hat R) \pa^\mu \Big] \phi \nn
&& = \int d^4 x \sqrt{\hat g}\, \frac{a}{(4\pi)^2} \Big[ (\hat\square\phi)^2
 - 2 \hat R^{\mu\nu} \pa_\mu\phi \pa_\nu\phi +\frac{2}{3}\hat R (\pa_\mu\phi)^2 \Big],
\label{quadratic}
\eea
and it is intriguing that for this choice the interaction terms in $\phi$ in Eq.~\eqref{intLiouville} drop out
and the theory is free (up to gravity interaction).
This action is precisely the conformally invariant one for a scalar field, as we see in eq.~\eqref{sacation21}
in Appendix~\ref{confs}.
So our Liouville action is expressed as
\bea
S_{4D, L}[g,\phi] =\lim_{D \to 4} \left( S_c[g e^\phi]-S_c[g] \right),
\label{consL}
\eea
where
\bea
S_c[g]=\int d^D x \sqrt{g}\ \frac{1}{(4\pi)^2}\Big(
\frac{c}{\epsilon}\, C_{\mu\nu\rho\s}^2 -\frac{a}{\epsilon}R_{GB}^2
 + \frac{a}{9} R^2 \Big).
\label{Sc}
\eea
Collecting all terms, our anomaly or Liouville action is
\bea
S_{4D, L}[\hg, \phi] \hs{-2}&=&\hs{-2} \int d^4 x \sqrt{\hg}\ \frac{1}{(4\pi)^2}\Big[
-c \phi C_{\mu\nu\rho\s}^2 + a \Big( \phi R_{GB}^2 - 2\hat G^{\mu\nu}\pa_\mu\phi \pa_\nu \phi
% + \frac19 \hat R^2
+(\hat\square\phi)^2 \nn
&& \hs{30} -\frac23 \phi \hat\square\hat R-\frac13 \hat R(\pa_\mu\phi)^2 \Big) \Big].
\label{final_liouville}
\eea

It is quite remarkable that the requirement that we should have consistent conformal anomaly leads to
the result that the Liouville mode becomes free field (up to gravity interactions), as in two-dimensional
Liouville theory.\footnote{
We can also confirm that this can be achieved if we use $R_{\mu\nu}^2$ for the local term but factor 3 difference.
From the formula~\eqref{q2}, we have
\bea
&& \hs{-10} \int d^4 x \sqrt{g}\, R_{\mu\nu}{}^2
= \int d^4 x \sqrt{\hg} \Big[
\hat R_{\mu\nu}{}^2
- 2 \hat R_{\mu\nu} \hat\nabla^\mu\hat\nabla^\nu\phi
+ \hat R_{\mu\nu} \pa^\mu\phi \pa^\nu\phi
- \hat R \hat\square \phi
-\hat R (\pa_\mu \phi)^2 \nn
&& \hs{10}+ (\hat\nabla_\mu\hat\nabla_\nu\phi)^2
+ 2 (\hat\square\phi)^2
- (\hat\nabla_\mu\hat\nabla_\nu\phi) \pa^\mu \phi \pa^\nu\phi
+\frac{5}{2} \hat\square\phi(\pa_\mu\phi)^2
+ \frac{3}{4} (\pa_\mu\phi)^2(\pa_\nu\phi)^2 \Big]\nn
&& \hs{-5} = \int d^4 x \sqrt{\hg} \Big[
\hat R_{\mu\nu}{}^2
- 2 \phi \hat\square \hat R + 3 (\hat\square\phi)^2 - \hat R (\pa_\mu \phi)^2
+ 3 \hat\square\phi(\pa_\mu\phi)^2
+ \frac{3}{4} (\pa_\mu\phi)^2(\pa_\nu\phi)^2 \Big].
\eea
It is then easy to show that the Liouville mode becomes free and we can obtain consistent anomaly
if we add this term with the factor $e/3$.
One can also check using \eqref{q1} that $R_{\mu\nu\rho\s}^2$ term also gives the same terms
as $R_{\mu\nu}^2$ and so we could use this. However $R^2$ is singled out since it can be introduced
as a gauge fixing term.}
Other choice of $e$ is not allowed since it cannot give consistent trace anomaly, and furthermore
it would give interacting Liouville mode.

Before beginning the quantization of Liouville theory, let us first confirm how the Liouville
action~\eqref{final_liouville} transforms under conformal transformations at the classical level.
As is evident from its construction~\eqref{consL}, we easily obtain the addition theorem
\bea
S_{4D, L}[g, \phi_1+\phi_2] =S_{4D, L}[ge^{\phi_1}, \phi_2]+S_{4D, L}[g, \phi_1].
\label{addL}
\eea
By substituting $\phi_1=\sigma, \phi_2=\phi-\sigma$ to this equation and moving terms properly,
we find how the Liouville action transforms under the conformal transformation at the classical level:
\bea
S_{4D, L}[ge^\sigma, \phi-\sigma] -S_{4D, L}[g,\sigma]=-S_{4D, L}[g, \sigma].
\label{transL}
\eea
Here the conformal transformation is defined by
\bea
g \mapsto ge^\sigma, \phi \mapsto \phi-\sigma .
\label{confphi}
\eea

As in the two-dimensional case \eqref{2DcL}, the anomaly coefficients of the Liouville
field~\eqref{final_liouville} are expressed as the sum of classical and quantum contributions.
\bea
c_{4D,L}=-\frac{1}{30}- c',  \nn
a_{4D,L}=-\frac{7}{180}- a'.
\label{4Dac}
\eea
Here, $-c'$ and $-a'$ are classical contributions coming from \eqref{transL}, and
the quantum one-loop anomaly from \eqref{quadratic} is given in~\eqref{anomalyf} with $x$ and $y$ in
Eq.~\eqref{xy}:
\bea
\frac{1}{(4\pi)^2} \left[ - \frac{1}{30} (\hat C_{\mu\nu\rho\s}^2-\hat\square\hat R)
 +\frac{7}{180}\hat R_{GB}^2 \right].
\eea

With these preparations in place, we can discuss the quantization of noncritical conformal gravity.
Our starting point is the following classically conformal theory:
\bea
S[g, \varphi^{\text{m}}]=S^{\text{m}}[g, \varphi^{\text{m}}]
+\int d^4 x \sqrt{g} \frac{1}{\lambda}\, C_{\mu\nu\rho\s}^2,
\label{origS}
\eea
where $S^{\text{m}}[g, \varphi^{\text{m}}]$ is the action for conformal matter
fields $\varphi^{\text{m}}$ with anomaly
coefficients $c^{\text{m}}$ and $a^{\text{m}}$.
We consider the noncritical case, that is, where \eqref{4Dcritical} is not satisfied.
We then separate the conformal modes by decomposing $g$ as
\bea
g_{\mu\nu}=\hg_{\mu\nu}e^\phi.
\label{decg}
\eea
Here the degrees of freedom of the conformal mode of $\hg$ are suppressed by the
gauge fixing of its conformal symmetry. Thus \eqref{origS} is equivalent to
\bea
S[\hg, \phi, \varphi^{\text{m}}]=\Big( S^{\text{m}}[g, \varphi^{\text{m}}]
+\int d^4 x \sqrt{g} \frac{1}{\lambda}\, C_{\mu\nu\rho\s}^2 \Big)_{g=\hg e^\phi}
+\int d^4 x \sqrt{\hg}\Big(\frac{1}{2\alpha} \hat{R}^2+b\square_{\hg}c \Big),
\label{decS}
\eea
If we ignore the conformal anomalies, the contributions to the partition function
of the first term in \eqref{decS}
does not depend on $\phi$ because the action is conformally invariant.
Taking the anomalies into account, or considering the counter terms of the form
\eqref{Sc}, we have
\bea
S[\hg, \phi, \varphi^{\text{m}}]=S^{\text{m}}[\hg, \varphi^{\text{m}}]
+\int d^4 x \sqrt{\hg} \frac{1}{\lambda}\, \hat{C}_{\mu\nu\rho\s}^2
+S_{4D,L}[\hg,\phi]
+\int d^4 x \sqrt{\hg}\Big(\frac{1}{2\alpha} \hat{R}^2+b\square_{\hg}c \Big).
\label{withL}
\eea
As in the two-dimensional case, the Liouville action can be regarded as the Jacobian for the change of
measure in the path integral:
\bea
\mathcal{D}_g g \,\mathcal{D}_g \phi \,\mathcal{D}_g \varphi^{\text{m}}
=\mathcal{D}_\hg \hg \,\mathcal{D}_\hg \phi \,\mathcal{D}_\hg \varphi^{\text{m}}
\, e^{-S_{4D,L}[\hg,\phi]},
\label{measure}
\eea
where $\mathcal{D}_g \varphi$ stands for the path measure for the field $\varphi$ defined on
the back ground metric $g_{\mu\nu}$.
As with the two-dimensional case, the parameters appearing in the Liouville action are difficult to determine
a priori because they undergo renormalization. However, they can be determined from the overall consistency as follows.

After rewriting $\hg$ to $g$ in \eqref{withL}, we have the total action:
\bea
S_{\text{total}}[g, \phi, \varphi^{\text{m}}]=S^{\text{m}}[g, \varphi^{\text{m}}]
+\int d^4 x \sqrt{g} \Big( \frac{1}{\lambda}\, C_{\mu\nu\rho\s}^2
+\frac{1}{2\alpha}\, R^2+b\square c \Big)
+S_{4D,L}[g,\phi],
\label{St}
\eea
where the last term $S_{4D,L}[g,\phi]$ contains two undetermined coefficients $a$ and $c$
(see \eqref{final_liouville}).
In fact we can determine them by the requirement
that $S_{\text{total}}$ have no anomaly for the BRS transformation of the conformal transformation, \eqref{brst1},
\eqref{brstc}, \eqref{brstb} and
\bea
\delta_B \phi=-\delta\lambda \, c.
\label{brstphi}
\eea
This is nothing other than the entire system, including the Liouville field and the ghost fields for fixing
the gauge of the conformal transformation, being invariant under the conformal transformation:
\bea
c_{\rm total}=c^{\text{conf}} +c^{\text{m}}+c_{4D,L} =0, \nn
a_{\rm total}=a^{\text{conf}} +a^{\text{m}}+a_{4D,L} =0.
\label{catotal}
\eea
Substituting \eqref{acconf}, \eqref{acm}, and \eqref{4Dac}, we have
\bea
\frac{199}{30}+\frac{1}{120}(N_S + 6N_F +12 N_V)- \frac{1}{30} - c' =0, \nn
\frac{87}{20}+\frac{1}{360}(N_S + 11N_F +62 N_V)-\frac{7}{180} - a' =0.
\eea

If this condition is satisfied, the BRS transformation is anomaly free at least at the 1-loop level and possesses
nilpotency.\footnote{Note that $\alpha$ in \eqref{St} is arbitrary as long as BRS symmetry is anomaly free.}
On the other hand, when considering higher-order corrections, the beta functions become more complex, and
it is not self-evident whether $c_{\rm total}$ and $a_{\rm total}$ can remain zero without receiving renormalization.

\subsection{Comments on Riegert proposal}

On the other hand, Riegert proposed the four-dimensional Liouville action as follows~\cite{R}.

Recall that the scalar curvature transforms as \eqref{scalart} in Appendix \ref{conft}.
We can calculate
\bea
\sqrt{g}\, \square R \hs{-2}&=&\hs{-2} \pa_\mu(\sqrt{g}\, g^{\mu\nu} \pa_\nu R) \nn
&=&\hs{-2} \sqrt{\hg}\, e^{(D-4)\phi/2} \Big[\hat\square \hat R - \frac{D-4}{2}\hat R(\pa_\mu \phi)^2
- \hat R\,\hat\square\phi +\frac{D-6}{2}\pa^\mu \hat R\, \pa_\mu\phi \nn
&& -\frac{(D-1)(D-2)}{2} \hat R^{\mu\nu}\pa_\mu \phi \, \pa_\nu \phi+(D-1)(\hat\square\phi)^2\nn
&& -\frac{(D-1)(D-2)(D-6)}{4} (\hat\nabla^\mu\pa^\nu\phi) \pa_\mu\phi \pa_\nu\phi +\frac{(D-1)(3D-10)}{4}
 \hat\square\phi(\pa_\mu\phi)^2 \nn
&& +\frac{(D-1)(D-2)(D-4)}{8}(\pa_\mu\phi)^2 (\pa_\nu\phi)^2 -(D-1) (D-4)\pa_\mu \hat\square\phi \pa^\mu\phi \nn
&& -(D-1) \hat\square^2 \phi -\frac{(D-1)(D-2)}{2}(\hat\nabla_\mu\hat\nabla_\nu\phi)^2 \Big].
\eea
Together with the results in \eqref{q1} --\eqref{q3} or \eqref{gbconforig}, we get
\bea
\sqrt{g} \Big[ R_{\rm GB}^2 -\frac{2}{3}\square R\Big] \hs{-2}&=&\hs{-2}
 \sqrt{\hg}\, e^{(D-4)\phi/2} \Big[\hat R_{\rm GB}^2
-\frac{2}{3} \hat\square \hat R +\frac{2(D-1)}{3} \hat\square^2 \phi - \frac13 (D-6) \pa^\mu\hat R \pa_\mu \phi \nn
&&\hs{-5}
-\frac{2}{3} (3D-10)\hat R \hat\square \phi
+ 4(D-3)\hat R^{\mu\nu} \hat\nabla_\mu \hat\nabla_\nu \phi
-\frac{(D-4)(3D-11)}{6}\hat R(\pa_\mu\phi)^2
\nn &&\hs{-5}
+\frac{(D-4)(D-5)}{3} \hat R^{\mu\nu} \pa_\mu\phi \pa_\nu \phi
-\frac{2}{3} (D-2)(D-4) (\hat\nabla_\mu \hat\nabla_\nu \phi)^2
\nn &&\hs{-5}
+\frac{(D-4)(3D-5)}{3} (\hat\square\phi)^2
+\frac{(D-2)(D+3)(D-4)}{6}(\hat\nabla^\mu\hat\nabla^\nu\phi)\pa_\mu\phi\, \pa_\nu\phi
\nn &&\hs{-5}
+\frac{(D-1)(D-2)(D-4)(3D-13)}{48} \{(\pa_\mu\phi)^2 \}^2
+\frac23 (D-1)(D-4)\pa^\mu\hat\square\phi \pa_\mu\phi
\nn &&\hs{-5}
-\frac16 (D-4)(3D^2-15D+16) \hat\square\phi (\pa_\mu\phi)^2 \Big].
\label{addrc}
\eea
If we set $D=4$, there remains only linear terms in $\phi$.
\bea
\sqrt{g} \Big[ R_{\rm GB}^2 -\frac{2}{3}\square R\Big] \hs{-2}&=&\hs{-2}
\sqrt{\hg} \Big[\hat R_{\rm GB}^2
-\frac{2}{3} \hat\square \hat R +2 \hat\square^2 \phi +\frac{2}{3} \pa^\mu\hat R \pa_\mu \phi
-\frac43 \hat R \hat\square\phi +4 \hat R^{\mu\nu} \hat\nabla_\mu\hat\nabla_\nu \phi \Big].\qquad\quad
\eea
It is an interesting observation that there remain only linear terms in $\phi$ in this combination~\cite{R}.
It was then proposed that these terms multiplied by $\phi/2$ give the anomaly action and is widely used since
then~\cite{AM}. However this does not follow the procedure to integrate the trace anomaly and it is not clear
why this has anything to do with the four-dimensional Liouville theory.
Moreover these terms are total derivatives, and there is no rationale why we can get the anomaly action just by
multiplying $\phi$ to these terms.
%{\red In fact, in order to to give the correctly normalized quadratic action in this proposal, we have to
%multiply Eq.~\eqref{addrc} by $\phi/2$ and then the normalization of the linear term differs from our Liouville
%action~\eqref{final_liouville} and the contribution at the classical level is not obtained correctly.}
On the other hand, it is clear that \eqref{final_liouville} gives the anomaly under the shift of
the conformal mode $\phi$ in the first order in $\phi$.

According to the general wisdom from two-dimensional quantum gravity, we should rather consider the counterterm
\bea
-\frac{1}{\e} \int d^D x \sqrt{g} \big[ R_{\rm GB}^2 -\frac{2}{3} \square R\big],
\eea
and make the Weyl transformation $g_{\mu\nu} \to e^\phi \hg_{\mu\nu}$ to extract the finite term
in the limit of $D\to 4$. This produces
\bea
\int d^4 x&&\hs{-6} \sqrt{\hg} \Big[\phi \Big\{ \hat R_{\rm GB}^2 -\frac23 \hat\square \hat R
+2 \hat\square^2 \phi  +\frac23 \pa^\mu\hat R\, \pa_\mu \phi -\frac43 \hat R \hat\square\phi
+4 \hat R^{\mu\nu}\hat\nabla_\mu\hat\nabla_\nu \phi \Big\} + \frac43 \hat\square^2 \phi
-\frac23 \pa^\mu\hat R \pa_\mu\phi
\nn &&
-4 \hat R\hat\square\phi
+ 8 \hat R^{\mu\nu} \hat\nabla_\mu \hat\nabla_\nu \phi - \frac{1}{3}\hat R(\pa_\mu\phi)^2
-\frac{2}{3} \hat R^{\mu\nu} \pa_\mu\phi \pa_\nu \phi
-\frac{8}{3}(\hat\nabla_\mu \hat\nabla_\nu \phi)^2
+\frac{14}{3} (\hat\square\phi)^2
\nn &&
+\frac{14}{3}(\hat\nabla^\mu\pa^\nu\phi)\pa_\mu\phi\, \pa_\nu\phi
-\frac{1}{4} \{(\pa_\mu\phi)^2 \}^2 -4(\hat\square \phi)^2
-\frac43 \hat\square\phi (\pa_\mu\phi)^2 \Big].\qquad
\label{rie}
\eea
The above proposal is just to keep the curly bracket terms, but this overlooks the remaining terms.
It is not clear how this can be justified.
Indeed it is easy to check that \eqref{rie} reduces to \eqref{gbcr} upon partial integration.
This is to be expected because the additional term in \eqref{addrc} is a total derivative.

\section{Summary}
\label{summary}

In this paper, we have first recalled the puzzle associated the beta functions for general quadratic curvature
theory and the conformal gravity, and show that the puzzle is resolved by considering the partial gauge fixing of
the conformal gauge symmetry with the BRST transformation. Extending the formulation with the BRST symmetry
to the noncrtical conformal gravity, we have proposed that the four-dimensional Liouville theory should be
given by Eq.~\eqref{final_liouville}, which is derived by deliberately adding the finite $R^2$ term~\eqref{combi}
such that the resulting action satisfies the requirement of the consistency of the conformal anomaly
and becomes a free quadratic action in the conformal mode, similar to two-dimensional Liuoville theory.
We have also given the condition that the BRST symmetry is
anomaly free, such that the quantum noncritical conformal gravity is consistent.
Finally, we have contrasted this approach with the widely-used Riegert proposal~\cite{R} for the four-dimensional
Liouville action, showing that our formulation is a more robust, consistent one derived through
the quantization procedure.

\section*{Acknowledgments}

H.K. is partially supported by the Ministry of Science and Technology, R.O.C.
(MOST 111-2811-M-002-016), and by National Taiwan University.
H.K. also thanks Prof. Shin-Nan Yang and his family for their kind support through the Chin-Yu chair professorship.
N.O. would like to thank Ningbo University, where this work was completed, for their kind hospitality and support.

\appendix

\section{Conformal transformation of curvature terms}
\label{conft}

If we make the Weyl transformation
\bea
g_{\mu\nu} = e^{\s} \tg_{\mu\nu},
\eea
the curvature tensors in $D$ dimensions transforms as
\bea
&& R_{\mu\nu} = \tilde R_{\mu\nu} - \frac{D-2}{2} \tilde\nabla_\mu \tilde\nabla_\nu \s
-\frac12 \tg_{\mu\nu} \tilde\square \s+\frac{D-2}{4} [\pa_\mu \s \pa_\nu \s
-\tg_{\mu\nu}(\pa_\a\s)^2], \nn
&& R = e^{-\s} \Big[ \tilde R-(D-1)\tilde\square\s-\frac{(D-1)(D-2)}{4}(\pa_\mu\s)^2 \Big].
\label{scalart}
\eea
A tilde indicates that the quantity is evaluated on the tilded metric $\tg$ and the indices are
raised, lowered and contracted by it.

The curvature square transforms as
\bea
\label{q1}
\sqrt{g}\, R^\mu{}_{\nu\a\b}{}^2
\!\! &=& \!\! \sqrt{\tg}\, e^{(D-4)\s/2} \Big[
\tilde R^\mu{}_{\nu\a\b}{}^2
- 4 \tilde R_{\mu\nu} \tilde\nabla^\mu \tilde\nabla^\nu \s + 2 \tilde R_{\mu\nu} \pa^\mu\s \pa^\nu\s
- \tilde R (\pa_\mu \s)^2  \nn
&& + (D-2) (\tilde\nabla_\mu\tilde\nabla_\nu \s)^2 + (\tilde\square\s)^2
- (D-2) (\tilde\nabla_\mu\tilde\nabla_\nu\s)\pa^\mu \s \pa^\nu\s
+ (D-2) \tilde\square \s(\pa_\nu\s)^2 \nn
&& + \frac{(D-1)(D-2)}{8} \{(\pa_\mu\s)^2\}^2 \Big],
\eea
\bea
\label{q2}
\sqrt{g}\, R_{\mu\nu}{}^2
\!\! &=& \!\! \sqrt{\tg}\, e^{(D-4)\s/2} \Big[
\tilde R_{\mu\nu}{}^2
- (D-2) \tilde R_{\mu\nu} \tilde\nabla^\mu\tilde\nabla^\nu\s
- \tilde R \tilde\square \s
+ \frac{D-2}{2} \tilde R_{\mu\nu} \pa^\mu\s \pa^\nu\s
 \nn
&& -\frac{D-2}{2}\tilde R (\pa_\mu \s)^2 + \frac{(D-2)^2}{4} (\tilde\nabla_\mu\tilde\nabla_\nu\s)^2
+ \frac{3D-4}{4} (\tilde\square\s)^2 \nn
&& - \frac{(D-2)^2}{4} (\tilde\nabla_\mu\tilde\nabla_\nu\s) \pa^\mu \s \pa^\nu\s
+\frac{(D-2)(2D-3)}{4} \tilde\square\s(\pa_\mu\s)^2 \nn
&& + \frac{(D-1)(D-2)^2}{16} (\pa_\mu\s)^2(\pa_\nu\s)^2 \Big],
\eea
\bea
\label{q3}
\sqrt{g}\, R^2
\!\! &=& \!\! \sqrt{\tg}\, e^{(D-4)\s/2} \Big[
\tilde R^2 -2(D-1) \tilde R \tilde\square\s
- \frac{(D-1)(D-2)}{2} \tilde R (\pa_\mu \s)^2 + (D-1)^2 (\tilde\square\s)^2 \nn
&& + \frac{(D-1)^2 (D-2)}{2} \tilde\square\s(\pa_\nu\s)^2
+ \frac{(D-1)^2 (D-2)^2}{16} \{(\pa_\mu\s)^2 \}^2 \Big],
\eea
The covariant derivatives on the rhs are all constructed by tilded metric $\tg_{\mu\nu}$.

\section{Conformally invariant scalar fields}
\label{confs}

Consider the usual kinetic term for a scalar field
\bea
\int d^4 x \sqrt{g}\ \frac12 g^{\mu\nu} \pa_\mu \vp \pa_\nu \vp.
\label{sacation1}
\eea
We consider the conformal transformation
\bea
g_{\mu\nu} = e^\s \tg_{\mu\nu}.
\eea
To make the action~\eqref{sacation1} invariant under this transformation, we should transform the scalar field as
\bea
\vp=e^{-\s/2} \tilde \vp.
\eea
Using the transformation property~\eqref{scalart}, we then find that the combination
\bea
\int d^4 x \sqrt{g}\ \frac12 \left[ g^{\mu\nu} \pa_\mu \vp \pa_\nu \vp +\frac{1}{6} R \vp^2 \right],
\label{sacation11}
\eea
is invariant under the conformal transformation.

Let us consider conformally invariant theory with higher derivative
\bea
\int d^4 x \sqrt{g}\ \frac12 (\square \vp)^2.
\label{sacation2}
\eea
To make this invariant under the transformation, we must transform the scalar field as
\bea
\vp=\tilde\vp.
\eea
Again using the transformation property~\eqref{scalart}, we find that the combination
\bea
\int d^4 x \sqrt{g}\ \frac12 \left[ (\square \vp)^2 -2 R^{\mu\nu}\pa_\mu\vp \pa_\nu \vp
+ \frac{2}{3} R g^{\mu\nu} \pa_\mu\vp \pa_\nu\vp \right],
\label{sacation21}
\eea
is invariant under the conformal transformation.

On the other hand, if we just take the d'Alembertian operator square $\square^2$, according to the formula
in Ref.~\cite{Gu}, it gives the anomaly
\bea
\frac{1}{(4\pi)^2} \left[ \frac{1}{120} \tilde C_{\mu\nu\rho\s}^2+\frac{1}{30}\tilde\square\tilde R
 -\frac{1}{360}\tilde R_{GB}^2 +\frac{1}{72} \tilde R^2 \right].
\eea
The coefficients of $\tilde C_{\mu\nu\rho\s}^2$ and $\tilde R_{GB}^2$ just correspond to those of two scalar
matter fields.
If we consider only normal kinetic term $-\square$, it gives
\bea
\frac{1}{(4\pi)^2} \left[ \frac{1}{120} \tilde C_{\mu\nu\rho\s}^2+\frac{1}{30}\tilde\square\tilde R
 -\frac{1}{360}\tilde R_{GB}^2 +\frac{1}{72} \tilde R^2 \right].
\eea
Surprizing enough, this also gives the same contribution as $\square^2$.
For the conformally invariant operator $-\square+\frac16 R$, it gives
\bea
\frac{1}{(4\pi)^2} \left[ \frac{1}{120} \tilde C_{\mu\nu\rho\s}^2+\frac{1}{180}\tilde\square\tilde R
 -\frac{1}{360}\tilde R_{GB}^2 \right].
\eea

\section{Consistency of the trace anomaly}
\label{consistency}

The general form of the quantum trace anomaly in four dimensions is~\cite{DDI}
\bea
\langle T_\mu^\mu \rangle &=& - \frac{2}{\sqrt{g}} g_{\mu\nu}
 \frac{\d S_{\rm eff}}{\d g_{\mu\nu}}(g_{\mu\nu}=e^{\phi} \hg_{\mu\nu}) \nn
&=& - \frac{\d S_{\rm eff}}{\d {\phi}} \nn
&\sim& R^2 +A R_{\mu\nu} R^{\mu\nu} + B R_{\mu\nu\la\rho} R^{\mu\nu\la\rho}.
\eea
up to overall factor. The consistency of the trace anomaly requires that
\bea
\d^2 S_{\rm eff}=0.
\label{cons1}
\eea
From the transformation property given in Appendix~\ref{conft}, we see under infinitesimal transformation $\d\s$,
in four dimensions,
\bea
\d^2 S_{\rm eff} = \int d^4 x \sqrt{g} \Big[ -6 R\, \square \d\phi
- A ( 2 R^{\mu\nu}\, \nabla_\mu \nabla_\nu \d\phi + R\, \square \d\phi)
- 4 B R^{\mu\nu} \nabla_\mu\nabla_\nu \d\phi \Big]\wedge \d\phi.
\label{conf1}
\eea
Here and in what follows, a tilde indicates that the quantity is evaluated on
the hatted metric $\hg$ and the indices are raised, lowered and contracted by it.
Making the partial integration to those terms with Ricci tensor, and using the Bianchi identity
\bea
\nabla_\mu R^{\mu\nu} =\frac12 \nabla^\nu R,
\eea
and making back the partial integration, \eqref{conf1} gives
\bea
\d^2 S_{\rm eff} = - \int d^4 \sqrt{g}\, 2(3+A+B) R\, \square \d\phi \wedge \d\phi \Big.
\label{cons2}
\eea
Namely the consistency gives
\bea
A+B+3=0.
\label{sol1}
\eea
Two independent solutions are
\bea
A=-4,\qquad
B=1,
\eea
which is precisely the GB term~\eqref{GBt}, and
\bea
A=-6, \qquad
B=3
\eea
which corresponds to the Weyl tensor square~\eqref{confg}.
To summarize, the trace anomaly is restricted to GB term and Weyl tensor squared by the consistency.
The important point is that there is no $R^2$ term.

The partition function is given by
\bea
Z[g e^\phi] =Z[g] e^{S_{\rm eff}[\phi,g]}.
\eea
When the consistency condition is satisfied, we have
\bea
Z[g e^\phi e^{\phi'}] &=& Z[g e^\phi] e^{-S_{\rm eff}[\phi',g e^\phi]} \nn
&=& Z[g] e^{-S_{\rm eff}[\phi,g] - S_{\rm eff}[\phi',g e^\phi]} \nn
&=& Z[g] e^{-S_{\rm eff}[\phi+\phi',g]}.
\eea
This means that
\bea
S_{\rm eff}[\phi+\phi',g]=S_{\rm eff}[\phi,g]+S_{\rm eff}[\phi',g e^\phi].
\eea

\end{document}